\documentstyle[twocolumn,aps,epsf]{revtex}
\begin{document}

\title{Diffusive foam wetting process in microgravity}

\author{H.Caps, M.-L.Chevalier, H.Decauwer, G.Soyez, M.Ausloos
and N.Vandewalle}

\address{GRASP, Institut de Physique B5, Universit\'e de Li\`ege, \\
B-4000 Li\`ege, Belgium.}

\address{~\parbox{14cm}{\rm\medskip
We report the experimental study of aqueous foam wetting in microgravity.
The liquid fraction $\ell$ along the bubble edges is measured and is found
to be a relevant dynamical parameter during the capillary process. The
penetration of the liquid in the foam, the foam inflation, and the rigidity
loss are shown all to obey strict diffusion processes.\\
\date{\today}}
\\
PACS: {82.70.Rr, 83.70.Hq}
}

\maketitle

\narrowtext

Foams are paradigms of disordered cellular systems. Bubbles composing foams
are indeed characterized by a wide variety of side numbers and faces areas
\cite{weaire}. The complexity of the foam can only be described by
statistical averages. Among the physical properties of interest, one can
cite the topological rearrangements \cite{rivier}, the cascades of popping
bubbles \cite{prlnico,jeff}, the rigidity loss transition \cite{rigidity},
etc...

In aqueous foams, a fundamental process is the drainage \cite{drainage}
which is due to the competition between gravity forces and the capillary
pressure in channels separating adjacent bubbles. The drainage-capillary
effects imply that the top of the foam becomes dry while the bottom of the
foam remains wet. A dry foam is composed of polyhedral bubbles meeting on
thin edges, while wet foams are composed of spherical bubbles which can
sometimes move freely \cite{rigidity}.

We report here the experimental study of foam wetting in microgravity. The
aims of the present letter are {\it(i)} to report the behavior of a foam in
microgravity and, {\it(ii)} to characterize the wetting of such a foam.
Physical mechanisms will be emphasized and characterized.

Microgravity experiments were held during a parabolic flight campaign
organized by the European Space Agency (ESA). About 30 parabolas have been
dedicated to this experiment. Parabolic flights allow for $20\, s$ in
microgravity with an average acceleration less than $a=0g \pm 0.05 g$. Each
parabola is composed of three parts. The {\it pull up} which is a
hypergravity phase ($a \simeq 2g$), when the plane is inclined at
$37^{\circ}$. The {\it microgravity} is established at the top of the
parabolic trajectory. During the {\it pull out}, i.e. the end part of the
parabola, the vertical acceleration is again $a \simeq 2g$.

The experimental procedure was the following. A soap-water mixture was
inserted in a vertical Hele-Shaw (HS) cell. The soap was mainly composed of
dodecylsulfate. The HS cells were closed parallelipedic vessels constituted
of 2 pieces of plexiglass ($20 \times 20 $ cm$^2$) distant of $0.2$ cm from
each other. This distance has been judicioulsy chosen in order to form only
one layer of bubbles, i.e. a two-dimensionnal foam. Before each parabola,
the HS cell was vigorously shaken for creating the foam. The HS cell was
placed vertically in a cage fixed to the plane for enhancing the drainage
before the microgravity phase. During the flights, a CCD camera recorded the
evolution of the foam. Image treatment and analysis have been later performed in
order to characterize the bubble edges and the liquid motion in the foam.

\begin{figure}[H]
\begin{center}
\centerline{\epsfxsize=7cm
\epsffile{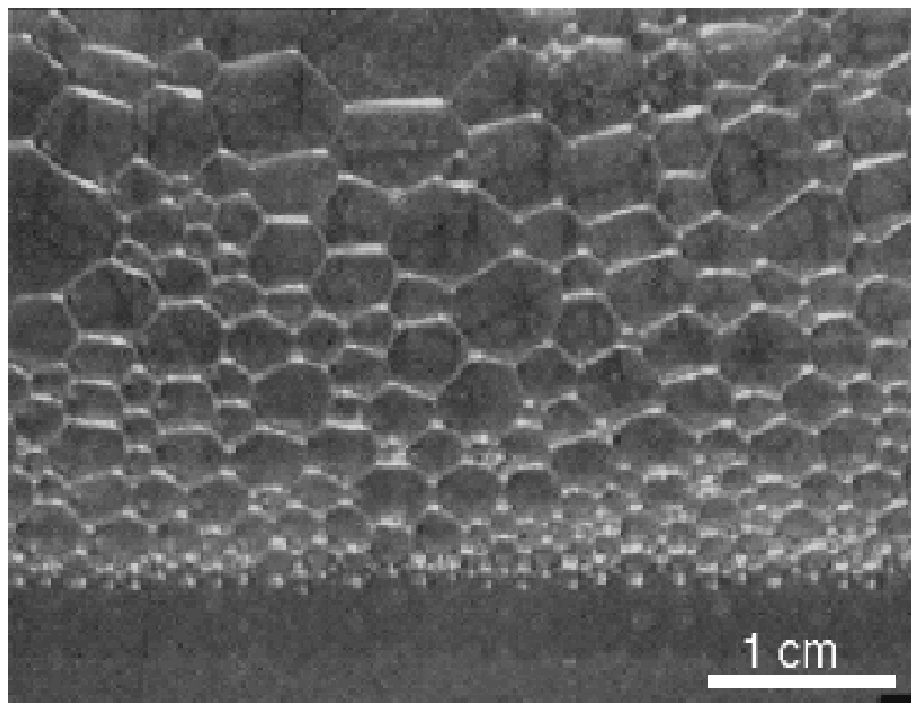}}
\centerline{\epsfxsize=7cm
\epsffile{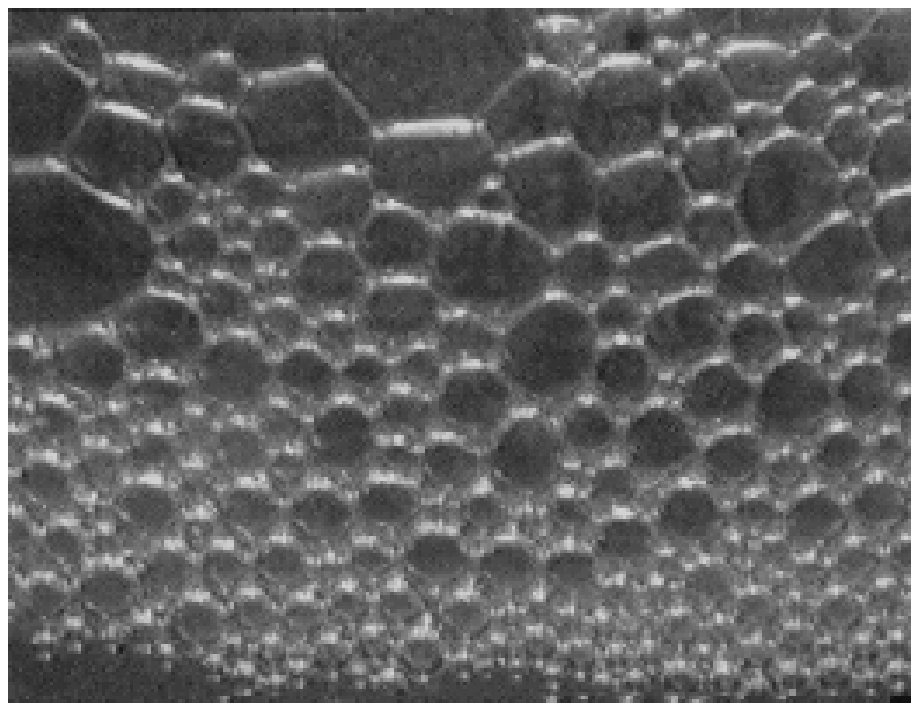}}
\vskip 0.2true cm
\caption{{\it (top)} Image of the foam during the hypergravity phase. The
bottom is composed of small `wet' bubbles while the top is composed of
large `dry' bubbles. {\it (bottom)} The same foam after $10$ seconds of
microgravity. All bubbles become spherical and bubble walls
thicken.}\label{hs}
\end{center}
\end{figure}

Figure \ref{hs} presents the foam during the hyper- and micro-gravity phases respectively. During the hypergravity phase, the bottom of the foam above the liquid is
composed of small ``wet'' bubbles while the top is composed of large polygonal ``dry'' bubbles as seen on the top picture of Figure 1. Some bubble
motion due to the plane vibrations is seen at the bottom of the foam. The
moves concern the small bubbles only. When the microgravity is established, the
situation changes drastically: the liquid invades the foam from below such that the average thickness of all bubble edges increases as seen in the lowest part of Figure \ref{hs}. The bubbles become more rounded and the rigidity of the foam is weakened,
allowing bubbles to slip on others and to move freely due to the airplane
vibrations. It should be noticed that small bubbles become rounded first.
The smallest ones are also dragged by the rising liquid towards the top of the
foam (see the central part of the bottom picture). Moreover, the front separating wet and dry phases is well seen to
propagate from bottom to top on the video records. Because of the liquid
invasion in the foam, the distance between adjacent bubbles grows and some
bubbles move down to the bottom of the HS cell. In other words, the foam
inflates. When the microgravity phase ends, an acceleration of about $2g$
leads to a fast and global drainage of the foam. The foam returns to the
initially dry situation quite rapidly, as in the top picture.

In order to quantify the wetting of the foam, the thickness of the bubble
edges has been measured by image analysis. The video record of each
parabola has been decomposed in a series of successive images at a rate
of 10 frames per second. Each image has been numerically modified for
enhancing the bubble edges (bright parts of Figure 1). On any horizontal
line situated at a vertical position $h$ on the images, the fraction of
bubble edges $\ell$ is measured. This corresponds to the liquid fraction at
that position. The parameter $\ell$ is given in units of the image width
such that $0<\ell<1$. A large value of $\ell$ corresponds to a wet foam,
while a small value of $\ell$ is the signature of a dry foam. One should
note that the origin of $h$ has been judiciously chosen such that the
bottom of the foam corresponds to $h=0$ at the end of the pull-up. We have analyzed more than 2500 images taken during 30 parabolas.

\begin{figure}[H]
\begin{center}
\centerline{\epsfxsize=8cm \epsffile{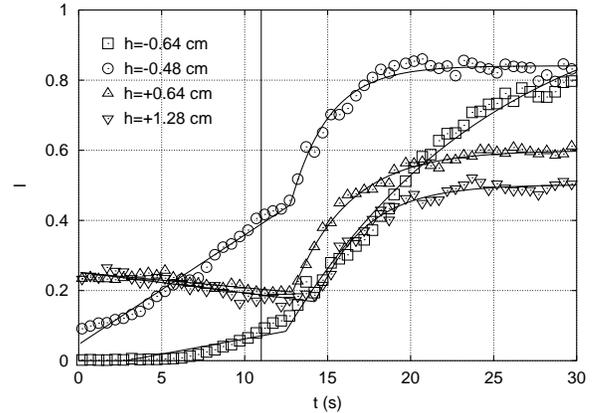}} \vskip 0.2true cm
\caption{Typical evolution of the liquid fraction $\ell$ as a function of
time $t$ for 4 different vertical positions $h$. Fits using
Eq.(\ref{eq_evol}) are shown. The vertical line corresponds to the
beginning of the microgravity phase, $t_m = 11$ s in this example. A break
of $\ell(t)$ is clearly observed at $t_0 > t_m$.}\label{l_evol}
\end{center}
\end{figure}

Figure \ref{l_evol} presents the typical evolution of the liquid fraction
$\ell$ as a function of time $t$ for 4 different vertical positions
$h$. Each dot corresponds to an average over 5 measurements, i.e. 5
parabolas. Only the pull-up phase and the microgravity phase are
illustrated on Figure 2. The beginning of the microgravity phase at time
$t_m$ is emphasized by a vertical line. All curves exhibit a break of
$\ell$ at $t_0$ after $t_m$.

The features of $\ell$ should be interpreted differently for $h>0$ and
$h<0$. Consider first the case $h>0$, e.g. $h=0.64$ cm and $h=1.28$ cm. During the hypergravity phase, the
foam is rigid and bubble edges are very thin. A small value of $\ell$ ($\approx 0.2$) is
seen in Figure 2 to be slightly decreasing with time, due to the acceleration phase. After the microgravity phase begins, a rapid growth of
$\ell$ is observed. This corresponds to the wetting of the foam, more
precisely the invasion of the liquid along bubble edges. The liquid
fraction saturates after some time. The dynamics of liquid invasion can
thus be extracted from $h>0$ measurements.

Consider now the $h<0$ curves of Figure \ref{l_evol},e.g. $h=-0.48$ cm and $h=-0.64$ cm. During the
hypergravity phase, only a few round bubbles are moving at the bottom of
the foam due to plane vibrations. This implies $\ell \ne 0$ on average even for
$h<0$. As the microgravity phase begins at $t_m$, the liquid invades the
foam which inflates since the bubbles are allowed to move towards the bottom
of the HS. A rapid growth of $\ell$ is observed for $t>t_m$ which
corresponds to the foam inflation. The liquid fraction $\ell$ is seen to
saturate about 10 seconds after $t_m$. Using the $h<0$ data, we can thus
study the foam inflation dynamics.

We thus see that the evolution of the inter bubble liquid fraction $\ell$ is a relevant
parameter in order to characterize the foam evolution. Considering that
$\ell$ saturates during the microgravity phase, we have assumed the
empirical law
\begin{equation}\label{eq_evol}
\ell\ =\ \left\{
\begin{array}{ll}
a+b\ (t-t_0)& \text{if $t<t_0$} \\
a+c \left(1- \exp\left(-(t-t_0)/\tau\right) \right)& \text{elsewhere}
\end{array}
\right.
\end{equation}
where $a$, $b$, $c$, $\tau$ and $t_0$ are 5 free fitting parameters at each height $h$. The
relevant physical parameters for our study are: the time $t_0$ at which the
liquid fraction becomes to grow for a given height $h$ and the
characteristic time $\tau$ of wetting. Both parameters will be examined
separately. Fits are shown in Figure \ref{l_evol}.

\begin{figure}[H]
\begin{center}
\centerline{\epsfxsize=8cm \epsffile{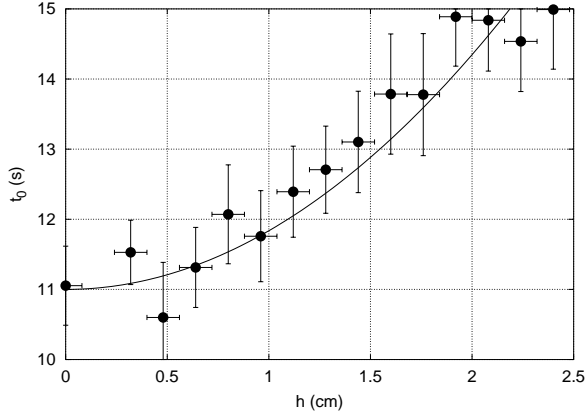}
} \vskip 0.2true cm
\caption{The time $t_0$ needed for the liquid to reach a height $h>0$. A
fit using Eq.(\ref{eq_h_t0_above}) is also shown.}\label{h_t0_above}
\end{center}
\end{figure}

\begin{figure}[H]
\begin{center}
\centerline{\epsfxsize=8cm \epsffile{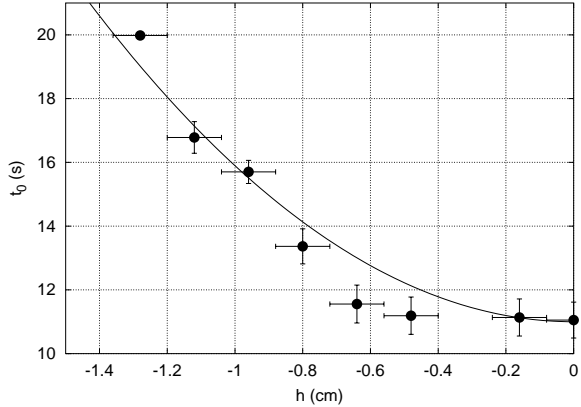}
} \vskip 0.2true cm
\caption{The time $t_0$ needed for the bubbles to fall down to a position $h<0$.
A fit using Eq.(\ref{eq_h_t0_under}) is also shown.}\label{h_t0_under}
\end{center}
\end{figure}

The parameter $t_0$ can be different from $t_m$ since there is a time delay
needed for the liquid to reach the vertical position $h>0$ or the foam
inflation for $h<0$. Figure \ref{h_t0_above} presents the time $t_0$ needed
to the liquid to reach the vertical position $h>0$. We have fitted the
results by a general power law, and have found a power exponent close to 2.
Thus, the wet front position behaves like
\begin{equation}\label{eq_h_t0_above}
h = \sqrt{D_w (t_0-t_m)} \, \, .
\end{equation}The liquid rise (wetting) in the initially dry foam is
clearly a diffusive process with a coefficient $D_w = 1.19 \pm 0.07$
cm$^2$/s. In addition to the liquid propagation, the foam inflation has
been observed. The dynamics of this process is captured by the parameter
$t_0$ for $h<0$ and is illustrated in Figure \ref{h_t0_under}. The foam
inflation behaves like
\begin{equation}\label{eq_h_t0_under}
- h = \sqrt{D_i (t_0-t_m)} \, \, .
\end{equation}The foam inflation is also a diffusive process with a
coefficient $D_i = 0.21 \pm 0.02$ cm$^2$/s. By comparing both coefficients,
one observes that $D_w > D_i$. Indeed, the motion of bubbles needed for the
foam inflation is a slow two-dimensionnal process with respect to the one-dimensionnal capillary rise of
liquid. In short, the foam wets faster than it inflates.

\begin{figure}[H]
\begin{center}
\centerline{\epsfxsize=8cm \epsffile{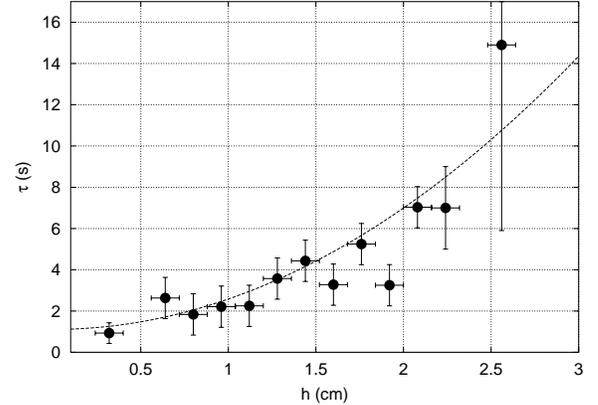}
} \vskip 0.2true cm
\caption{Characteritic rigidity loss time $\tau$ as a function of the height $h>0$. Fit
using Eq.(\ref{eq_h_tau_above}) is also illustrated.}\label{tau_above}
\end{center}
\end{figure}

\begin{figure}[H]
\begin{center}
\centerline{\epsfxsize=8cm \epsffile{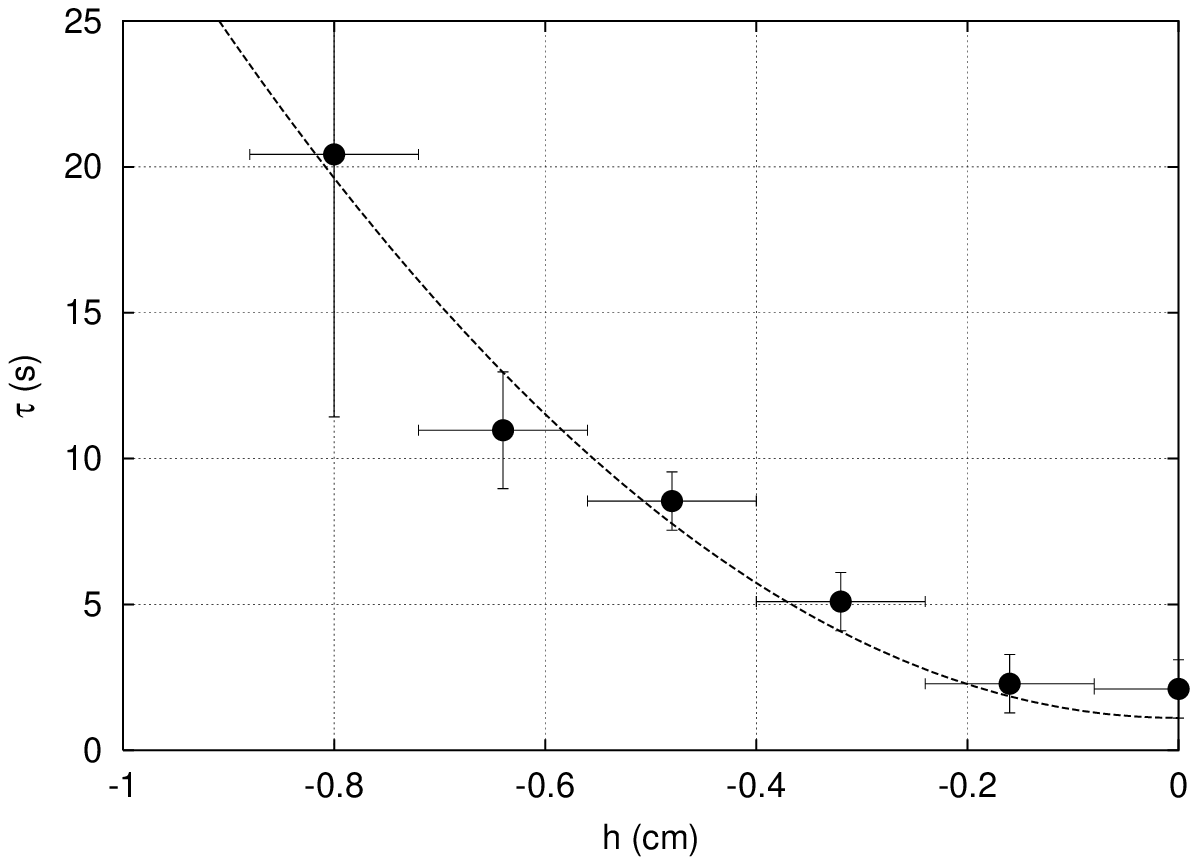}
} \vskip 0.2true cm
\caption{Characteritic horizontal diffusion time $\tau$ as a function of the height $h<0$. Fit
using Eq.(\ref{eq_h_tau_under}) is also illustrated.}\label{tau_under}
\end{center}
\end{figure}

Once adjacent bubbles are wetted by the rising liquid, they start to move apart because the bubble separation increases. This process lasts until the foam rigidity is lost. Then, the
bubbles can move independently. The rigidity loss can be captured by the
parameter $\tau$ for $h>0$. In Figure \ref{tau_above}, we report the
measurement of $\tau$ as a function of the vertical position $h>0$. We have
found that a quadratic expression fits the data. One has
\begin{equation}\label{eq_h_tau_above}
h = \sqrt{D_r \tau} \, ,
\end{equation} meaning that the bubbles take a long time to separate at the
bottom of the HS cell. Again, a diffusive behavior is found with a
coefficient $D_r$. The kinetics of rigidity loss is close to the one of
wetting but twice lower. Figure \ref{tau_under} shows the measurement of
$\tau$ as a function of the vertical position $h<0$. This represents the
horizontal motion of bubbles at the bottom of the HS cells once the foam
inflates. This is also a diffusion process
\begin{equation}\label{eq_h_tau_under}
- h = \sqrt{ D_m \tau} \, ,
\end{equation} with a small coefficient $D_m$. Indeed, the motion of
bubbles in a wet phase is quite small.

In the case of drainage under gravity, an equation has been proposed by
Verbist and coworkers \cite{drainage} for describing a Poiseuille flow
along bubble edges. The drainage equation does not apply here because of
the microgravity ($g \approx 0$) and the foam inflation we observed. Except
for capillary forces at the beginning of the process, {\it (i)} no driving force pushes
the liquid in our case, and {\it (ii)} one sees (after some time) the individual motion of bubbles in
a liquid foam. Diffusive motion is thus the key process during foam
wetting, foam inflation and rigidity loss. The ``square root of time"
behaviors that we observe confirm this hypothesis. Table \ref{tab_D}
lists the values of the various diffusion coefficients encountered in the
present study.

\begin{table}
\centering
\begin{tabular}{lcc}
    & $t_0$  & $\tau$ \\
\hline
$h>0$ & $D_w=1.19 \pm 0.07$ cm$^2$/s & $D_r=0.68 \pm 0.07$ cm$^2$/s\\
$h<0$ & $D_i=0.21 \pm 0.02$ cm$^2$/s & $D_m=0.035 \pm 0.002$ cm$^2$/s\\
\end{tabular}
\vskip 11pt
\caption{The different diffusion coefficients measured in our experiment.
The liquid rise (wetting) is characterized by $D_w$. The foam inflation is
characterized by $D_i$. The rigidity loss is characterized by $D_r$. The
motion of individual bubbles in the wet phase is characterized by
$D_m$.}\label{tab_D}
\end{table}

In summary, we have studied experimentally the dynamics of foam wetting in
microgravity. We have confirmed that the liquid invasion behaves clearly as
a diffusive process.

\section*{Acknowledgements}

HC is financially supported by FRIA (Brussels, Belgium). GS is financially
supported by FNRS (Brussels, Belgium). The team thanks the European Space
Agency for the facility, and M. Valentiny for his recommandations.

\end{document}